  \documentclass[12pt]{article}
  \usepackage{graphicx}
  \advance\textheight by \topskip
\newcommand{\ba}{\begin{array}}
\newcommand{\ea}{\end{array}}
\newcommand{\bd}{\begin{displaymath}}
\newcommand{\ed}{\end{displaymath}}
\newcommand{\be}{\begin{equation}}
\newcommand{\ee}{\end{equation}}
\newcommand{\bea}{\begin{eqnarray}}
\newcommand{\eea}{\end{eqnarray}}

\def\q2 {q^2}

\def\lapp{\mathrel{\rlap{\raise.5ex\hbox{$<$}}
                    {\lower.5ex\hbox{$\sim$}}}}
\def\gapp{\mathrel{\rlap{\raise.5ex\hbox{$>$}}
                    {\lower.5ex\hbox{$\sim$}}}}

 \def\N10{\widetilde \chi_1^0}
                         \def\C1p{\widetilde \chi_1^+}
                         \def\C1m{\widetilde \chi_1^-}
                         \def\C1pm{\widetilde \chi_1^\pm}

\def\lsim{\buildrel{\scriptscriptstyle <}\over{\scriptscriptstyle\sim}}

\def\mET{E_T \hspace{-1.1em}/\;\:}

\def\beq{\begin{eqnarray}}
\def\enq{\end{eqnarray}}

\def\lsim{\:\raisebox{-0.5ex}{$\stackrel{\textstyle<}{\sim}$}\:}

\newcommand{\decay}[2]{
\begin{picture}(25,20)(-3,3)
\put(0,-20){\line(0,1){10}}
\put(0,-20){\vector(1,0){15}}
\put(0,0){\makebox(0,0)[lb]{\ensuremath{#1}}}
\put(25,-20){\makebox(0,0)[lc]{\ensuremath{#2}}}
\end{picture}}
\begin{document}
\begin{flushright}
{\large WUE-ITP-2004-039}\\
{\large SHEP-04-40}\\
\end{flushright}
\begin{flushleft}
\end{flushleft}
\begin{center}
{\Large\bf Probing the light neutral Higgs boson scenario 
of the CP-violating MSSM Higgs sector at the LHC }\\[15mm]
{\large Dilip Kumar Ghosh$^{a,}$\footnote{dghosh@physik.uni-wuerzburg.de}
and Stefano Moretti$^{b,}$\footnote{stefano@hep.phys.soton.ac.uk}}\\[4mm]
{\em $^{a}$ Institut f\"ur Theoretische Physik und Astrophysik, \\ 
Universit\"at W\"urzburg, D-97074, W\"urzburg, Germany}\\[2mm]
{\em $^{b}$ School of Physics \& Astronomy, University of
Southampton, \\
Highfield, Southampton SO17 1BJ, United Kingdom}\\[10mm]
\end{center}
\begin{abstract}
\noindent
In the CP-violating Minimal
Supersymmetric Standard Model (MSSM), for certain  values of 
the  CP-violating phases associated to the universal 
trilinear couplings $(A_t, A_b)$ and the gluino mass
$(M_{\tilde g})$, e.g., $\Phi_{\rm{CP}} = 60^\circ$ or $90^\circ$,
for $M_{H^+} \lsim 140 $ GeV
and $\tan \beta \sim 2-5 $, the lightest Higgs boson mass ($M_{H_1}$) is
$\lsim ~50 $ GeV. This mass interval is still allowed 
from standard LEP Higgs searches because of a strongly
suppressed $H_1ZZ$ 
coupling. However, in the same region of 
parameter space in which these two conditions occur, the $H_1 H^\mp W^\pm $ 
coupling is enhanced because the two mentioned 
sets of couplings satisfy a sum rule. 
In this paper we probe such a light
Higgs scenario at the Large Hadron Collider 
(LHC) by studying $H^\pm H_1$ associate production,
leading to a $4b + \ell^\pm + \mET $ signal. We show that the latter
is readily accessible at the CERN hadron collider,  upon the application
of suitable selection cuts against the Standard Model (SM) backgrounds. Our
parton level Monte Carlo (MC) analysis yields $\sim 15-45 $ signal events, 
 completely free of SM background, for
${\cal L} = 10-30~{\rm fb}^{-1}$ of accumulated luminosity, 
after taking into account the overall efficiency for tagging four 
$b$-jets. 

\end{abstract}

\vskip 1 true cm

\noindent

\newpage

The experimental observation of Higgs bosons and the determination of their
properties is crucial for the understanding of Electro-Weak Symmetry Breaking
(EWSB).
Thus, the search for Higgs bosons is one the major goals of the present 
collider Tevatron (Run II) and future ones as well, such as 
the forthcoming LHC and the planned International 
Linear Collider (ILC). In the SM, the Higgs boson mass is not 
predicted. Over the last few decades, many efforts have been put into 
detecting such a particle, but to no avail. 
From direct searches at LEP, a lower bound of $114$ GeV 
has been set on its mass \cite{Eidelman:2004wy,lep}. In the 
MSSM, with all real and CP-conserving
parameters, the lower limit on the lightest Higgs boson is $\sim 90$ GeV
\cite{susylim} for any $\tan\beta$. However, this bound is 
significantly lowered in an MSSM scenario with radiatively induced Higgs sector
CP-violation, as the latter in turn implies a suppressed $H_1ZZ$ coupling 
\cite{gunion}.

CP-violation in the Higgs sector is possible in multi-Higgs doublet models, 
such as a general 2-Higgs Doublet Model (2HDM) or indeed the MSSM. 
In the latter, it has been shown that, assuming universality of the 
gaugino masses $(M_i, i =1,2,3)$ at some high energy scale, the CP-violating
MSSM Higgs sector can be parametrised in terms of two independent phases: 
that of the Higgsino mass parameter (also called  $\mu$ term), i.e.,
 ${\rm Arg}(\mu)$, 
and that of the soft trilinear Supersymmetry (SUSY) 
breaking parameters, i.e., ${\rm Arg}(A_f) $, with $f=t,b$. 
The experimental
upper bounds on the Electric Dipole Moments (EDMs) of electrons and 
neutrons 
\cite {edm1, edm2} as well as of mercury atoms \cite{edm3} may pose severe  
constraints on these phases. 
However, these 
limits are highly model dependent. In particular, it has been shown that one 
could still have large CP-violating phases yielding relevant EDM values all 
satisfying current experimental bounds if any 
of these three possibilities is realised: $(a)$ the sfermions of the
first two generations are heavy, of the order of a few TeV \cite{edm4}; 
$(b)$  cancellations between different EDM contributions \cite{edm5}
take place; 
$(c)$ universality of the trilinear scalar couplings $A_f$ is dismissed
\cite{edm6,edm7}. However, only in the scenario with first and second 
generation 
sfermions much heavier than the third generation ones, the 
phase of $\mu$ can be large. Otherwise, it is  strongly constrained, as 
${\rm Arg}(\mu) \lsim 10^{-2}$.

In the MSSM, non-zero phases of $\mu$ and/or the trilinear 
scalar couplings $A_{f}$ can induce 
CP-violation at one-loop level 
in the Higgs sector even in presence of a CP-conserving 
tree-level scalar potential, through the
CP-violating interactions among 
Higgs bosons and heavy sfermions. This
one-loop corrected Higgs potential then generates non-zero off-diagonal 
 terms ${\cal M}^2_{SP}$ 
in the $3\times 3$ neutral Higgs boson mass-squared matrix ${\cal M}^2_{ij}$,
representing mixing
between  CP-even (or scalar, S) and CP-odd (or pseudoscalar, P) states 
\cite{cpv1}--\cite{cpv6}. These off-diagonal entries can approximately
be written as follows \cite{cpv2}:
\beq
{\cal M}^2_{\rm SP} \approx {\cal O}\left
( \frac{M^4_t  \mid \mu \mid \mid A_t \mid}{v^2 32 \pi^2 M^2_{\rm SUSY}}\right )
\sin \Phi_{\rm CP}  \left [6, \frac{\mid A_t \mid^2 }{M^2_{\rm SUSY}},
\frac{\mid \mu\mid^2}{\tan\beta M^2_{\rm SUSY}},
\frac{\sin 2\Phi_{\rm CP}\mid A_t\mid \mid\mu\mid }{\sin \Phi_{\rm CP}
M^2_{\rm SUSY}}\right ],
\enq
where $\Phi_{\rm CP} = {\rm Arg}(A_t)={\rm Arg}(\mu)$ and $v = 246$ GeV.  The mass scale
$M_{\rm SUSY}^2$ is typically defined to be $(m^2_{\tilde t_1} + m^2_{\tilde t_2})/2 $,
i.e., in terms of the two stop masses.
After diagonalising the $3\times 3$ symmetric Higgs mass-squared matrix
${\cal M}^2_{ij}$ by an orthogonal matrix $O$, the physical mass
eigenstates $H_1, H_2 $ and $H_3$ (in ascending order of mass) are
states of indefinite  CP-parity.  In this case, $M_{H^\pm}$ is the
most appropriate mass parameter to describe the MSSM Higgs-sector (in place
of $M_A$, often  used in the CP-conserving case).
\begin{figure}[hbt]
\begin{center}
 \includegraphics*[scale=0.6] {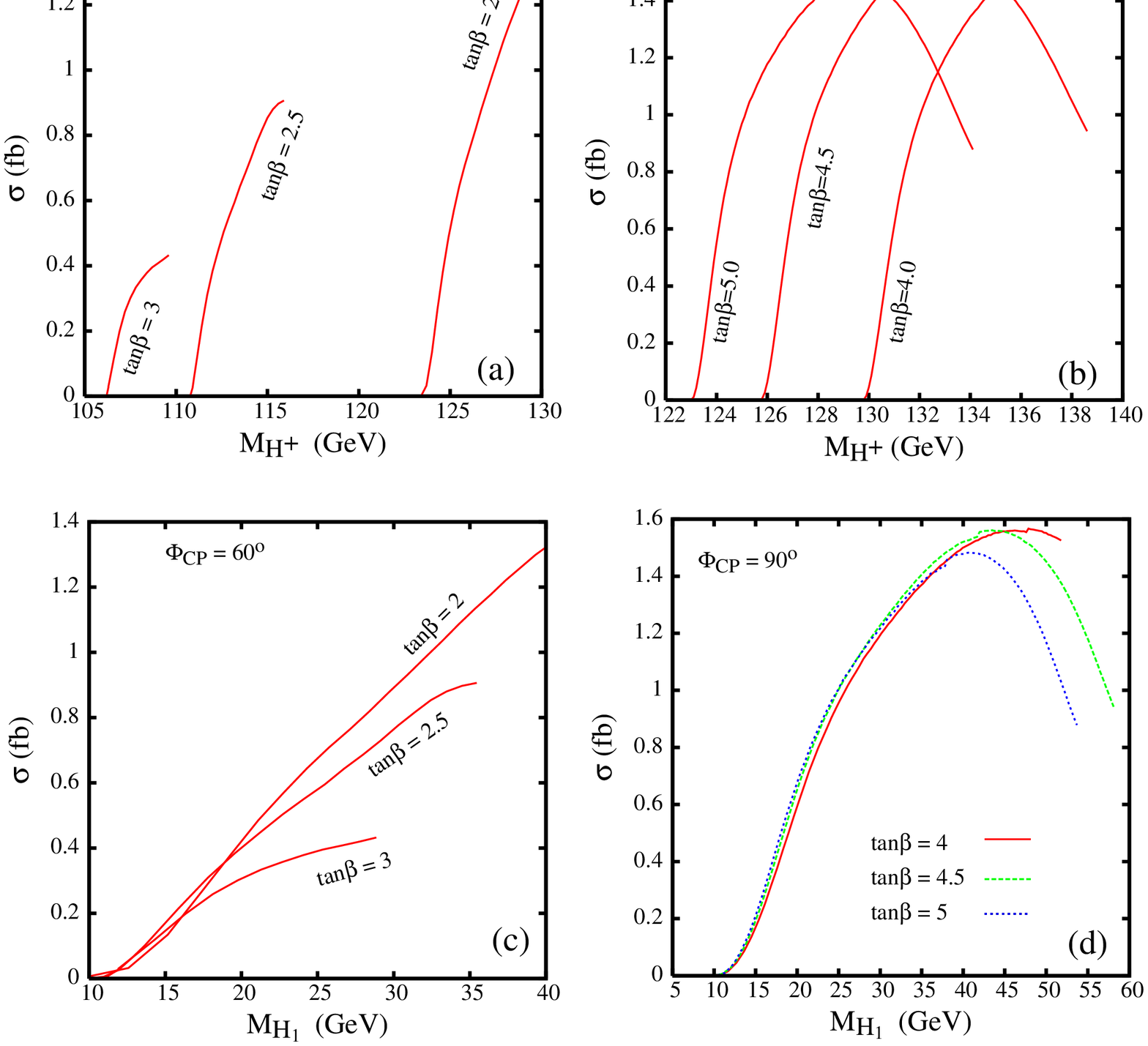}
\vspace*{-5.0cm}
\caption{Variation of the signal cross-section with $M_{H^\pm}$ [$(a),(b)$]
and $M_{H_1}$ [$(c), (d)$] for the two values of CP-violating phases 
$\Phi_{\rm CP} = 60^\circ $ [$(a),(c)$] and   
$\Phi_{\rm CP} = 90^\circ $ [$(b),(d)$]. The choices of $\tan\beta $ for
each CP-violating phase are shown in the figure.}
\end{center}
\label{fig1}
\end{figure}

\begin{figure}[hbt]
\begin{center}
\includegraphics*[scale=0.75] {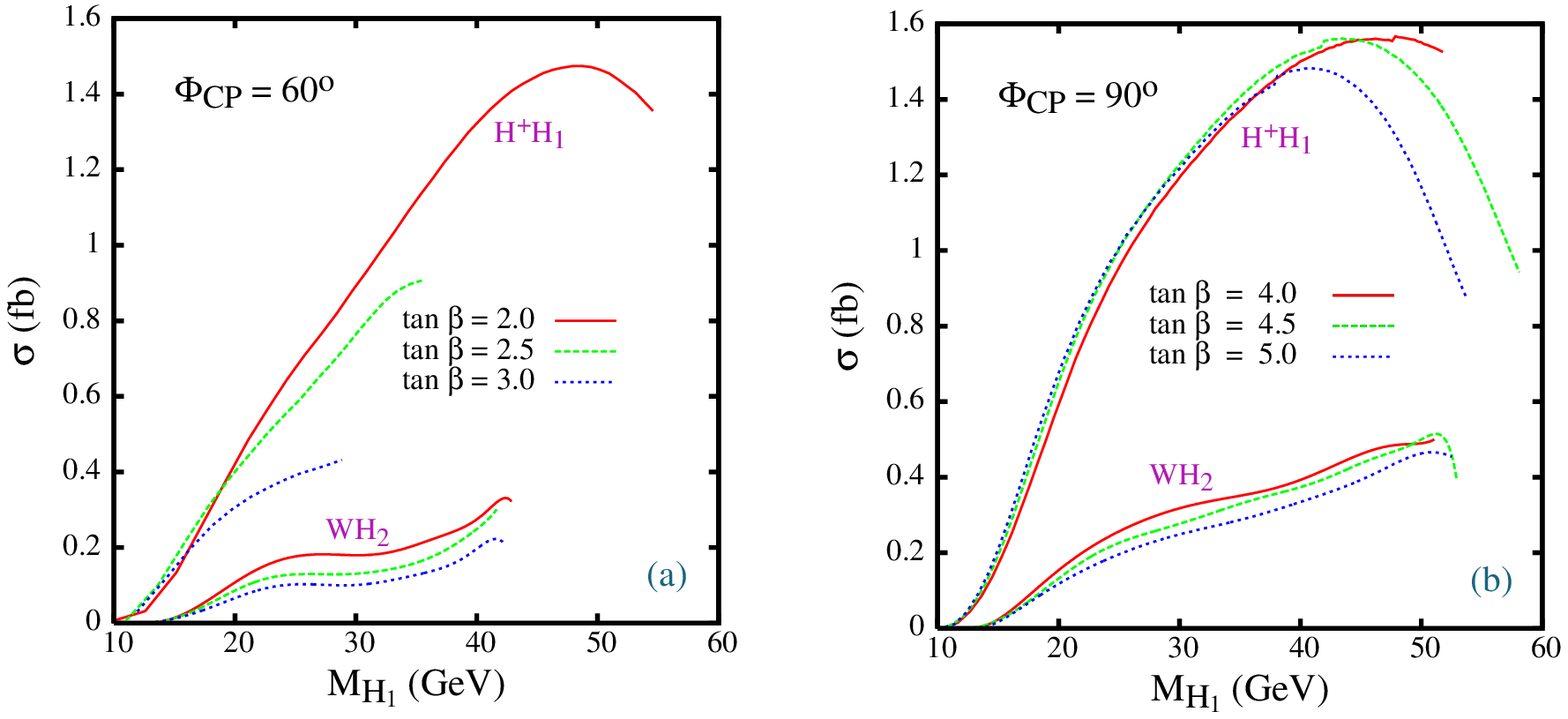}
\vspace*{-14.0cm}
\caption{A comparison between the production rates of the two 
processes $q\bar q' \to H_2 W^\pm\to H_1H_1W^\pm \to b\bar bb\bar b\ell\nu$ 
and  $q\bar q' \to H_1 H^\pm  \to H_1H_1W^\pm \to b\bar bb\bar b\ell\nu$,
for $\Phi_{\rm{CP}}=60^o$ $(a)$ and $\Phi_{\rm{CP}}=90^o$ $(b)$, for the three
choices of $\tan\beta $ shown in the figure.}
\end{center}
\label{fig_compare}
\end{figure}

The CP-violating phases can cause the Higgs couplings to fermions and gauge 
bosons to change significantly from their tree-level values, 
with dramatic consequences for MSSM Higgs phenomenology at present and 
future colliders \cite{cpv7}. Here, we start by 
presenting the form of the MSSM couplings $H_iVV$, $H_iH_jZ$ and 
$ H_iH^- W^+$, where $V=W^\pm,Z$, in presence of explicit 
CP-violation\footnote{Further details can be found in Ref.~\cite{cpv2}.}:
\beq
{\cal L}_{H_iVV}&=& g m_W\ \sum_{i=1}^3 g_{H_i VV}
\ [H_iW_\mu^+W^{-\mu} +\frac{1}{2 c_W^2} H_iZ_\mu Z^{\mu} ], \\
{\cal L}_{H_iH_j Z} &=& \frac{g}{2 c_W}\ \sum_{j>i=1}^3
g_{H_iH_jZ}\ ( H_i\, \!\!
\stackrel{\leftrightarrow}{\vspace{2pt}\partial}_{\!\mu} H_j )Z^\mu, \\
{\cal L}_{H_i H^- W^+}&=& \frac{g}{2 c_W}\ \sum_{i=1}^3
g_{H_iH^-W^+}\ ( H_i\, \!\!
\stackrel{\leftrightarrow}{\vspace{2pt}\partial}_{\!\mu} H^-)W^{+\mu},
\enq
where  
\beq
g_{H_iVV} &=& O_{1i}\cos\beta+ O_{2i}\sin\beta, \\
g_{H_iH_j Z} &= &O_{3i}(\cos\beta O_{2j}- \sin\beta O_{1j})
-(i\leftrightarrow j),\\
g_{H_iH^-W^+} &=& O_{2i}\cos\beta- O_{1i}\sin\beta + i O_{3i}.
\enq
These couplings obey the following sum rules:
\beq
\sum_{i=1}^3 g^2_{H_i VV} & = & 1,\\
g^2_{H_i VV} + \mid g_{H_i H^-W^+}\mid^2 & = & 1, \\
g_{H_k VV}  & = & \epsilon_{ijk} g_{H_i H_j Z}.
\end{eqnarray}
Hence, from the above formulae, one can see that -- if two of the 
$g_{H_i ZZ}$ are known -- then the whole set of couplings between neutral
Higgs bosons and  gauge
bosons is determined. It is also interesting to see from Eq. (9)  that -- in
the presence of large CP violating effects  with large scalar-pseudoscalar
mixing -- a suppressed $H_1 VV$ coupling means an enhanced strength of the
$H_1 H^-W^+$ vertex, as intimated already. This enhancement will indeed play a
significant role in our analysis. Equally important is the correlation 
between the mass of the charged Higgs state, $M_{H^\pm}$, and that of the
pseudoscalar state, $M_A$. In fact, a suppressed $H_1 VV$ coupling implies a
light pseudoscalar state which in turn leads to a light charged Higgs, in 
particular $M_{H^\pm} < m_t$ (the top quark mass).

From Eq.~(1),  it is clear that sizeable scalar-pseudoscalar mixing is
possible for a large value of the CP-violating phase $\Phi_{\rm CP}$ and/or of
$\mid \mu \mid $ and $\mid A_t \mid $. In this respect, the CP-violating benchmark 
scenario defined as CPX in \cite{cpv2}
provides a suitable choice of MSSM parameters which maximises CP-violating
effects and can then be used to study the most striking  phenomenological
manifestations of  
explicit CP-violation in the MSSM Higgs sector. Such a scenario is defined as followed:
\beq
M_{\tilde Q}= M_{\tilde t} = M_{\tilde b} = M_{\rm SUSY},\\
\mu=4 M_{\rm SUSY}, \qquad |A_t|=|A_b|= 2 M_{\rm SUSY},\\
{\rm Arg}(A_t)={\rm Arg}(A_b).
\enq
Recently, the OPAL Collaboration \cite{opal_collab} reported
their results for Higgs boson searches in a CP-violating MSSM Higgs
sector using the parameters defined in the CPX scenario 
and found that for the CP-violating phases 
$\Phi_{\rm CP}= 60^\circ$ and $90^\circ $ and certain values of
$M_{H^\pm}$ and $\tan\beta $, the lowest
mass limit on the neutral Higgs is very light -- at times even vanishing 
completely -- hence resulting in some low Higgs mass windows in the 
$M_{H^\pm}$--$\tan \beta$ 
plane which are still allowed by LEP data. In a nutshell, the reason for the
existence of such  regions is the fact
that  in the CPX scenario the lightest Higgs boson is almost CP-odd with highly suppressed 
couplings to $ZZ$ pairs.

The experimental analysis was done by adopting both CPSuperH~\cite{cpsuperh} and FeynHiggs 
2.0~\cite{Heinemeyer:2001qd}, as these are the two public codes available 
for the calculation of masses and mixing angles  
in the CP-violating MSSM Higgs sector. In reality, these two programs
give somewhat different results -- at least in the case of the
CPX scenario -- mainly due to  different 
approximations used in their calculations. To give  more conservative constraints, the  
experimentalists used the lower prediction of the two for the expected Higgs boson cross-sections. 
The constraints also depend sensitively on the mass of the top
quark used in the calculation~\cite{susylim}\footnote{They
were obtained for $m_t=175$ GeV, the value we adopt here.}. 
The results of ~\cite{susylim}, from
a combined analysis of all LEP data, provide exclusion
regions in the $M_{H_1}- \tan \beta $ plane for the following values of the 
SUSY parameters:
\beq
{\rm Arg}(A_t) = {\rm Arg}(A_b) = {\rm Arg}(M_{\tilde g})=\Phi_{\rm CP},\\
M_{\rm SUSY}= 0.5~{\rm TeV}, \qquad M_{\tilde g} = 1~ {\rm TeV},\\
M_{\tilde B} = M_{\tilde W}= 0.2~ {\rm TeV},\\
\Phi_{\rm CP} = 0^{\circ}, 30^{\circ}, 60^{\circ}, 90^{\circ},
\enq
where $M_{\tilde B}, M_{\tilde W}$ and $M_{\tilde g}$ 
represent the soft SUSY-breaking masses for bino, wino and gluino.


By combining the results of Higgs searches from ALEPH, DELPHI, L3 and OPAL,
the authors of Ref.~\cite{cpv8} also provided exclusion regions in the
$M_{H_1}- \tan \beta $ (as well as $M_{H^+}- \tan \beta $) plane for
the same  set of parameters. The exact shape of the exclusion regions 
may be somewhat different in their analyses, but they
all show that for certain values of CP-violating phases LEP cannot rule out
a light Higgs mass at low values of $\tan\beta$. This interesting situation 
roughly corresponds to 
$\tan \beta \sim 3.5-5, ~M_{H^\pm}\sim 125-140 $ GeV (yielding $ M_{H_1}
\stackrel{<}{{}_\sim} 50 $ GeV) and $\tan \beta \sim 2-3, ~M_{H^\pm}\sim 105-130
~{\rm{GeV}}$ (yielding $M_{H_1} \stackrel{<}{{}_\sim} 40 $ GeV), respectively. 
The authors of 
Ref.~\cite{cpv8} further showed that in the same regions the $H_1 t \bar t$
coupling is suppressed too.  Thus,  these two particular areas of the
parameter space can be probed neither at the Tevatron -- where  the
$W H_1$ associated production mode is the most promising one -- nor
at the LHC -- as the reduced $H_1t \bar t $ coupling suppresses
both the inclusive production mode $gg\to H_1$ and  the associated 
one $t \bar t H_1$. 

We have however found that, in the mentioned regions of the
$M_{H^+}- \tan \beta $ plane, the decay $H^\pm \to H_1 W^\pm $
has a very large $(\sim 100 \%) $  Branching Ratio (BR), thanks
to the discussed enhancement of the $H^\mp H_1 W^\pm $ coupling 
and the mass hierarchy $m_t>M_{H^\pm}\gg M_{H_1}$. This feature motivated
us to then study the possibility of probing such a light Higgs scenario in the 
CP-violating MSSM Higgs model through the process 
$p p \to H_1 H^\pm \to H_1 (H_1 W^\pm) \to b \bar b b \bar b \ell \nu $ 
at the LHC. Recently, in Ref.~\cite{self}, the authors have probed this 
light Higgs scenario through $t\bar t $ production at the LHC, where one of 
the top quarks decays into $bb\bar b W$, via the chain $t\to b H^\pm, 
H^\pm \to W^\pm H_1$ and $H_1 \to b \bar b $, leading to a 
$4b + jj + \ell^\pm +\mET$ signal. In our analysis, the signal will 
consist of up to four $b$-jets along with a hard lepton (electron or muon) 
and missing transverse energy $\mET$, according to the following
decay pattern: \\
$$
p p \rightarrow \decay{H^\pm~~~}{{{\decay{W^\pm}{\ell \nu }}~~~~~~~~
{\decay{H_1}{b~\bar b}}}} ~~~~~~~~\hspace{1cm} + \hspace{1cm} \decay{H_1}{ b~\bar b} + ~~~~X
\vspace{3cm}
$$
In short, such a production and decay mechanism should allow to probe the possible
existence of a light $H_1$ state  
in the above mentioned two interesting windows of the CP-violating MSSM in the CPX scenario.
The production mechanisms 
$p p \rightarrow H^\pm H_1$ and 
$p p \rightarrow H^\pm H^0$  
in the CP-violating MSSM and 2HDM, respectively, were
discussed in Ref.~\cite{andy}, where it was remarked upon the large 
$H^\pm \to H_1 W^\pm $ BR, yet no
phenomenological analyses were reported there. 

Our numerical results are obtained from a parton level MC
analysis, wherein partons are treated as jets. As acceptance and selection criteria we 
required: 
\begin{enumerate}
\item $\mid \eta\mid <2.5 $ for all jets and leptons, where $\eta$ denotes 
pseudo-rapidity;
\item $p_T^{b-{\rm jets}} > 15 $ GeV;
\item $p_T^{\ell}>10 $ GeV $(\ell = e, \mu )$;
\item $\mET > 20 $ GeV;
\item a minimum separation $\Delta R \equiv
\sqrt{ \left (\Delta \phi \right)^2 +
\left (\Delta \eta \right )^2} = 0.4$ between leptons and jets 
as well as each pair of jets; 
\item  reconstruction of the two light Higgs
bosons, by requiring four $b$'s in the event to be tagged as such\footnote{With 
no jet- and/or lepton-charge determination, though.}  
and that at least one out of the three possible double 
pairings of $b$-jets satisfies the following mass constraint:
\begin{equation} 
\label{eq:mhchi2}
\frac{(m_{b_1,b_2}-M_{H_1})^2+(m_{b_3,b_4}-M_{H_1})^2}{\sigma_m^2} < 2, 
\end{equation} 
where $\sigma_m= 0.12~ M_{H_1}$.
\end{enumerate}
(In enforcing the latter constraint, we implicitly assume that trial 
values for $M_{H_1}$ are attempted and the selection has hit on the right one.)
Notice that we impose Gaussian smearing on energies, 
with $\Delta E/E = 0.6/\sqrt{E}$
for jets and $\Delta E/E = 0.12/\sqrt{E}$ for leptons, to 
realistically emulate finite 
experimental resolution. All such cuts and smearing procedure have been applied
to any  process studied in this paper, alongside using
CTEQ(5L) \cite{CTEQ5} as Parton Distribution Functions (PDFs) taken at the
scale $Q^2=\hat s$ (same for the scale of $\alpha_{\rm S}$, where relevant).

In Figure~1 we show the variation of the signal 
cross-section (including the suppression due to a quadruple $b$-tagging 
efficiency, $\epsilon_b^4=(1/2)^4=1/16$)
with $M_{H^+}$ [$(a)$ and $(b) $] and $M_{H_1}$ [$(c)$ and $ (d)$] for the 
CP-violating phase choices $\Phi_{\rm CP} = 60^o$ [$(a)$ and $(c)$] 
and $90^\circ $[$(b)$ and $(d)$], respectively, 
for three values of $\tan\beta$. The choice of other MSSM parameters is 
defined through Eqs.~(14)--(16). We have used the 
CPSuperH program~\cite{cpsuperh} with $m_t = 175 $ GeV 
to calculate the masses, couplings and decay rates
of the relevant Higgs bosons and semi-analytical techniques  to
evaluate the hadro-production cross-section and decay rates. 
From Figure~1$(a)$ and $(b)$ one can see
that the signal cross-section has a peculiar dependence upon $M_{H^\pm}$.
This may seem counter-intuitive, as the light Higgs mass increases with
increasing $M_{H^+}$. However, it should be noticed that, 
at lower $M_{H_1}$ values,
the $b$-jets emerging from the Higgs decays are rather soft and close
to each other in phase space. As the light Higgs mass increases though,
$b$-jets become harder and also acquire
much larger angular separations, hence a kinematics satisfying the 
cuts mentioned above more often, counter-balancing the decline in
production rates due to larger $M_{H_1}$ values, ultimately
leading to an overall relative rise in the cross-section.
The final drop in the latter is due to phase space suppression
for the $H^\pm \to H_1 W^\pm$ decay. In this scenario the largest cross-section,
$\sim 1.36~{\rm fb} $, can be obtained for $\Phi_{\rm CP} = 60^\circ $, 
$\tan\beta = 2$ and $M_{H^+} \sim 130 $ GeV, which corresponds to 
$M_{H_1} \sim 40 $ GeV. For the CP violating phase $\Phi_{\rm CP} = 90^\circ $,
the largest cross-section, $\sim 1.5 $ fb,  can be obtained for
$\tan\beta = 4$  and $M_{H^+} \sim 139 $ GeV, which corresponds to 
$M_{H_1} \sim 50 $ GeV.  From Figure $1 (b)$ and $(d)$, it is interesting 
to notice that the cross-section is almost insensitive to
$\tan\beta $ at low values of $M_{H_1}$. 

The SM background cross-section 
arising from\footnote{Hereafter, $j$ labels both light ($u,d,s,c$) and heavy ($b$) quark jets.} 
\begin{enumerate}
\item QCD production of $gg \to b \bar b jj\ell \nu $;
\item Electro-Weak (EW) $q\bar q'\to ZZ W^\pm $ production, followed by the decay of each $Z$ into $b\bar b$ pairs 
and by electron/muon decays of the $W$-boson;
\item top-quark production and decay via $gg\to t\bar t\to b \bar b  jj \ell \nu$;
\end{enumerate}
is not shown, as it is negligible. In fact,
after applying the same cuts 1.--6. to both signal
and background processes and 
folding the 
cross-sections with the usual $b$-tagging efficiency ($\epsilon_b = 1/2$
per each $b$-jet) and 
the appropriate light-quark jet rejection factors 
(e.g., assuming $R_{u,d,s} = 1/50$ and $R_c = 1/25$) \cite{ATLAS_TDR},
we found that 
\begin{enumerate}
\item $ \sigma (gg \to b \bar b jj\ell \nu) \lsim 2.2 \times 10^{-3} $ fb; 
\item $\sigma (q\bar q'\to ZZ W^\pm \to b\bar b jj \ell \nu)
\lsim 4.0 \times 10^{-3} $ fb;
\item $\sigma (gg\to t\bar t\to b \bar b  jj \ell \nu) \lsim 2.9 \times 10^{-2}
$ fb;
\end{enumerate}
where we have taken into account all relevant $Z,W^\pm$ BRs 
and combinatorial factors.

In Ref.~\cite{cpv8}, the authors  
discussed $q\bar q' \to H_2 W^\pm $ as another possible probe of CP-violation
in the MSSM Higgs sector, wherein 
$W^\pm\to \ell\nu$ and $H_2\to H_1H_1\to b\bar bb\bar b$.
This mode can then lead to the same signature as the one we have been
considering. 
Hence, one should in principle worry about its numerical relevance 
as well as possible interference effects between the two channels. In practise
though, the $q\bar q' \to H_2 W^\pm \to H_1H_1W^\pm \to b\bar bb\bar b\ell\nu$ 
production and decay rates are much smaller than those for 
$q\bar q' \to H_1 H^\pm  \to H_1H_1W^\pm \to b\bar bb\bar b\ell\nu$. 
This is clearly confirmed by Figure~$2$. This is also the case for the
interference. Therefore, hereafter, we will neglect considering
the $q\bar q' \to H_2 W^\pm$ channel further.

\begin{figure}[hbt]
\begin{center}
\includegraphics*[scale=0.75] {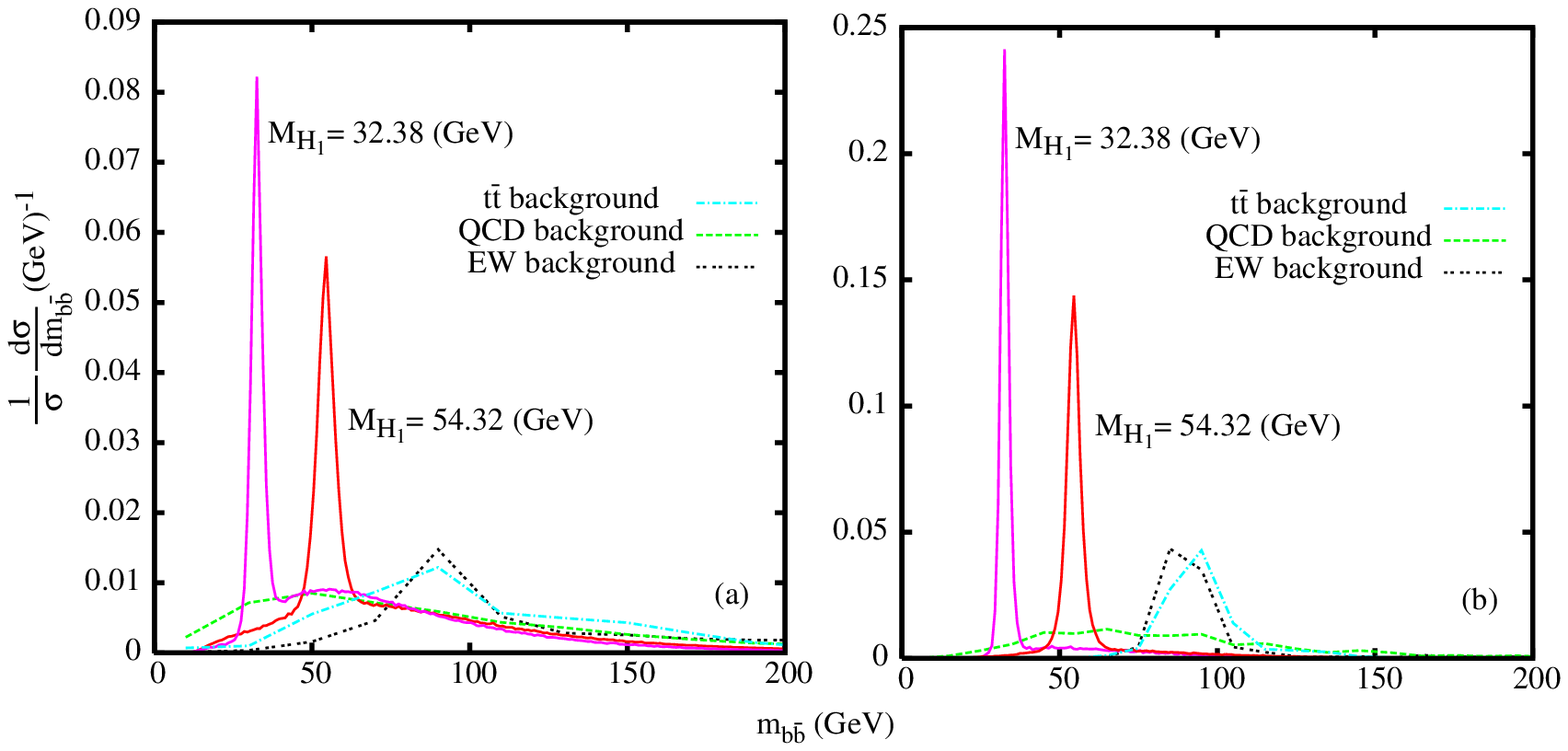}
\vspace*{-11.0truecm}
\caption{The `average' $(a)$ and `best reconstructed' $(b)$
$m_{b\bar b}$ distributions (as described
in the text) for the signal (with $M_{H_1}= 32.38$ and $54.32$ GeV, 
corresponding to $M_{H^\pm}=129.55$ and $136.97$ GeV, respectively, for 
$\tan\beta = 4.5$ and $\Phi_{\rm{CP}}=90^\circ$). The corresponding
SM model background distributions are also shown.}
\end{center}
\label{fig_masses}
\end{figure}


Note that signal events will be very striking due to the 
clustering of two pairs of $b\bar b$ invariant masses around 
$M_{H_1}$. This feature will of course not be present in the backgrounds,
so it can be used to enhance the former and suppress the latter.
One can attempt to reconstruct the light Higgs mass
in the following ways. Out of the $4b$ final state, one can simply plot
all possible (six) combinations of invariant masses $m_{b\bar b}$ (each
with identical weight). This leads to the 
signal and background `average'  $m_{b\bar b}$  distributions
appearing in Figure~$3(a)$. Alternatively, one can construct the three 
possible double pairings of $b\bar b$ invariant masses, then select the 
pairing giving 
the least difference between the two $m_{b\bar b}$  values, 
and the best reconstructed $H_1$ mass is the corresponding
mean value of that pair. This `best reconstructed' $m_{b\bar b}$ distribution
is shown in Figure~$3(b)$. 
By comparing the spectra in Figures~$3(a)$ and $3(b)$, it is
clear that the second procedure is more efficient than the first one
in increasing the signal-to-background rate. (Hence, we have exploited
it by adopting the mass constraint described in Eq.~(\ref{eq:mhchi2}).)
For the two discussed mass spectra, two sample values of the light
Higgs boson mass are assumed in the signal. Also note the normalisation 
to unity for all processes considered. 
Naturally, in producing these plots, we have refrained from applying
cut 6. above.
Figures~$3(a)$ and $3(b)$ clearly highlight the $H_1$ resonant peaks
for the signal and
the $Z$ ones for the EW background. As for the other two
background processes,
the QCD one reveals a typical Jacobian shape, owning to the
absence of heavy
particles decaying hadronically, while the $t\bar t$ one
displays $W^\pm$
resonance effects, when two light quarks are mistagged as
$b$-jets.
In all cases, it is worth noting that the Gaussian smearing
we have applied to
the momenta somewhat affect the actual locations of the
resonant peaks (where
relevant). Under any circumstances, it is clear from both
figures that the low mass
signal resonances are always located far away from the bulk
of all background events
appearing at high mass.

\begin{figure}[hbt]
\begin{center}
\includegraphics*[scale=0.75] {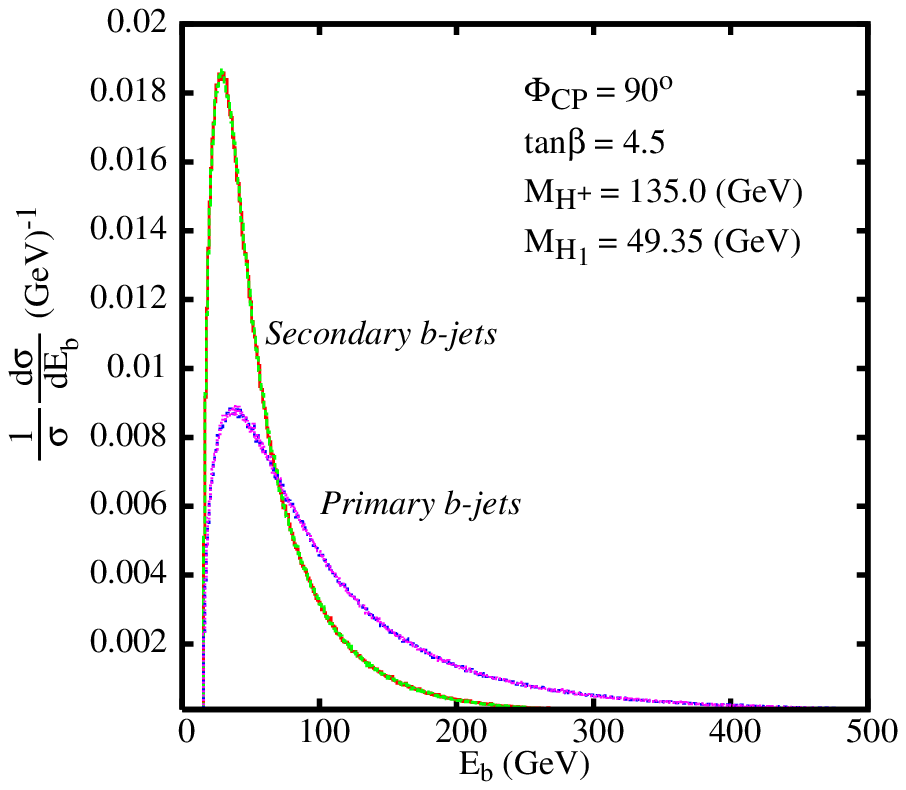}
\caption{The energy distributions of the two `primary' and two `secondary'
$b$-jets in the signal (as defined in the text), as obtained for the 
representative parameter choices shown inside the figure.} 
\end{center}
\label{figEb}
\end{figure}
The question now arises of whether it is also possible to reconstruct
the $H^\pm$ mass in the signal. Because of the presence of a neutrino
escaping detection in the $W^\pm\to \ell\nu$ decay, the actual 
charged Higgs resonance is not kinematically accessible. Besides, only
one of the two $b\bar b$ pairs selected via Eq.~(\ref{eq:mhchi2}) actually
comes from a $H^\pm\to H_1 W^\pm\to b\bar b W^\pm$ decay chain. In order
to obviate such potential problems, we have proceeded as follows. Firstly,
by running our MC for the signal, we have verified that the two (hereafter,
`primary') $b$'s emerging from 
the decay of the $H_1$ boson produced in the hard scattering in association
with the charged Higgs boson have normally a higher energy than the two
produced in the mentioned decay chain (hereafter, `secondary'). This is evident 
from Figure~$4$, obtained after all cuts 1.--6. have been enforced.
Hence, it makes sense to identify, between the two $b\bar b$ pairs selected
via Eq.~(\ref{eq:mhchi2}), the one with least total energy as the one
produced in the mentioned $H^\pm$ decay chain. Secondly, we have defined 
the transverse mass, $M_T$, constructed from, on the one hand,
the visible transverse momentum of the system formed by the 
`secondary' $b\bar b$ pair plus the lepton and, on the other hand, 
 the missing energy, i.e.,
\begin{equation}\label{mT} 
M_T = \sqrt { \left (p^b_T + p^{\bar b}_T + p^\ell_T + {p\!\!\!/}_T \right)^2
   -\left (\bf p^b_T + \bf p^{\bar b}_T + \bf p^\ell_T + \bf {p\!\!\!/}_T \right)^2 },
\end{equation}
where quantities in boldface refer to three-vectors.
Such a variable is sensitive, as clearly evident from Figure~5, to the
underlying charged Higgs boson mass and thus can be used within a MC
simulation to fit the latter.

\begin{figure}[hbt]
\hspace*{0.5truecm}\includegraphics*[scale=0.75] {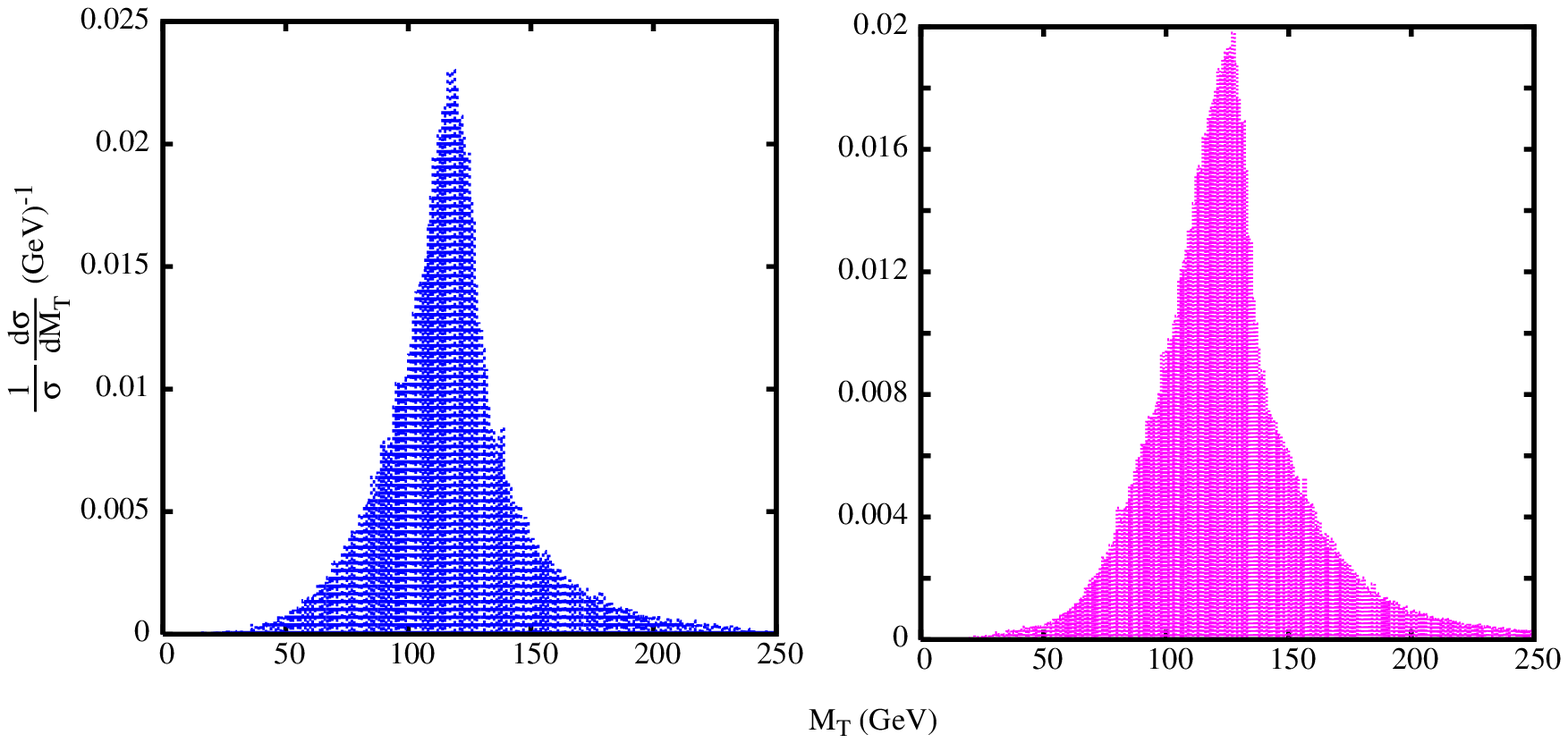}
\vspace*{-12.0cm}
\caption{The transverse mass distribution  
(as defined in the text) for the signal, for  
$M_{H^\pm}=129.55 $ GeV (left panel) and $M_{H^\pm}=135 $ GeV (right panel), 
$\tan\beta = 4.5$ and $\Phi_{\rm{CP}}=90^\circ$.}
\label{fig_mT}
\end{figure}
In summary, we have proved the feasibility of testing
at the LHC  light Higgs boson
windows in the so-called CPX scenario of the CP-violating
MSSM,  wherein a $H_1$ signal
might have been lost at LEP due to a strongly suppressed
$H_1 Z Z$ coupling. Specifically, we have concentrated our attention
upon the following two MSSM parameter space regions: (i)
$3.5 < \tan\beta  < 5,~M_{H_1} \lapp 50 $ GeV and (ii)
$2 < \tan \beta  < 3, ~M_{H_1} \lapp 40$ GeV, assuming
a common CP-violating
phase $\Phi_{\rm CP} = 90^{\circ}$ and $60^{\circ}$,
respectively. We have
found that, in the above mentioned parameter space areas,
a light charged Higgs boson ($M_{H^\pm} < m_t$) can
decay to $W^\pm H_1$ pairs with large a branching fraction
so that, combined with a sizeable $H^\pm H_1$ associate
production rate, the yield of the emerging signature $
4b+\ell^\pm + \mET $ is sufficient to isolate a
CP-violating signal. About $\sim 15-45 $
signal events, completely free of SM background, for ${\cal L}=10-30 $ 
fb$^{-1}$ of accumulated luminosity, after taking into account the
efficiency for tagging all four $b$-jets, can be isolated. Furthermore, we have also 
discussed the possibility of measuring the light Higgs boson mass as well
as the charged Higgs boson one. We expect that the parton level study
presented in this paper will encourage the CMS and ATLAS collaborations 
to carry out further investigations of the MSSM in presence of explicit 
CP-violation in the Higgs sector.

\section*{Acknowledgements}
The work of DKG is supported by the Bundesministerium f\"ur Bildung 
und Forschung Germany, grant 05HT1RDA/6. DKG would also like to thank
A. Datta and T. Binoth for discussions.


\end{document}